\documentclass[12pt,a4paper]{article}

\usepackage{epsfig}
\usepackage{amsmath,amsfonts,amssymb}
\usepackage{t1enc}
\usepackage{cite}

\providecommand{\openone}{\leavevmode\hbox{\small1\kern-3.8pt\normalsize1}}

\newcommand{\RE}{\text{Re}\,}
\newcommand{\IM}{\text{Im}\,}

\newcommand{\smn}{\sigma^{\mu \nu}}

\newcommand{\gm}{\gamma^\mu}

\newcommand{\pam}{\partial_\mu}

\newcommand{\DMD}{\overleftrightarrow{D^\mu}}
\newcommand{\Wmn}{W_{\mu \nu}}
\newcommand{\Wm}{W_\mu}

\newcommand{\Bmn}{B_{\mu \nu}}

\newcommand{\Dsl}{D\!\!\!\!\!\!\not\,\,\,}
\newcommand{\qsl}{q\!\!\!\!\!\not\,\,}
\newcommand{\psl}{p\!\!\!\!\!\not\,\,}

\newcommand{\ptsl}{p_t\!\!\!\!\!\!\!\not\,\,\,\,}
\newcommand{\pcsl}{p_c\!\!\!\!\!\!\!\not\,\,\,\,}

\newcommand{\vl}{V_L}
\newcommand{\vr}{V_R}
\newcommand{\gl}{g_L}
\newcommand{\gr}{g_R}
\newcommand{\xttl}{X_{tt}^L}
\newcommand{\xttr}{X_{tt}^R}
\newcommand{\dvz}{d_V^Z}
\newcommand{\daz}{d_A^Z}

\newcommand{\xctl}{X_{ct}^L}
\newcommand{\xctr}{X_{ct}^R}
\newcommand{\kl}{\kappa_{ct}^L}
\newcommand{\kr}{\kappa_{ct}^R}

\newcommand{\yttv}{Y_{t}^V}
\newcommand{\ytta}{Y_{t}^A}

\newcommand{\yctl}{\eta_{ct}^L}
\newcommand{\yctr}{\eta_{ct}^R}
\newcommand{\FL}{\omega_L}
\newcommand{\FR}{\omega_R}
\newcommand{\yqtl}{\eta_{qt}^L}
\newcommand{\yqtr}{\eta_{qt}^R}
\newcommand{\Fqtl}{\omega_{qt}^L}
\newcommand{\Fqtr}{\omega_{qt}^R}
\newcommand{\Fqtlc}{\omega_{qt}^{L*}}
\newcommand{\Fqtrc}{\omega_{qt}^{R*}}

\begin{document}

\begin{center}
\begin{Large}
{\bf A minimal set of top-Higgs anomalous couplings}
\end{Large}

\vspace{0.5cm}
J. A. Aguilar--Saavedra  \\[0.2cm] 
{\it Departamento de Física Teórica y del Cosmos and CAFPE, \\
Universidad de Granada, E-18071 Granada, Spain} \\[0.1cm]
\end{center}

\begin{abstract}
We use the equations of motion to simplify the general form of fermion-fermion-Higgs interactions generated by dimension-six gauge-invariant effective operators. After removing redundant operators it is found that the most general $H f_i f_j$ vertex for off-shell fermions $f_i$, $f_j$ and an off-shell Higgs boson only involves scalar and pseudo-scalar terms, without derivatives. Examples are presented for the $Htt$ and $Htq$ interactions, where $q=u,c$, giving the explicit expressions of the vertices in terms of gauge-invariant operators.
The new operator equalities obtained here also reduce the number of operators relevant for the $Ztq$ vertices, although the general form of these interactions is not simplified with respect to previous results.
\end{abstract}

\section{Introduction}

The discovery of the Higgs boson is one of the main goals and motivations of the Large Hadron Collider (LHC). If this particle is finally found, as it is widely expected, its discovery will constitute a major advance in our understanding of electroweak symmetry breaking. The study of its properties and couplings to the Standard Model (SM) particles will then be of the greatest importance to determine its nature and fully understand the symmetry breaking mechanism, and it may possibly provide signals of new physics beyond the SM.
On the other hand, the top quark also deserves special interest among the known fermions. Being the heaviest quark, effects of new physics on its couplings are expected to be larger than for other fermions, so it is hoped that deviations with respect to the SM predictions might be found with precise measurements of its couplings. Hence, in the case of the top-Higgs couplings the measurements have added relevance, also bearing in mind that the top itself might have a special role in electroweak symmetry breaking due to its large mass.

It is known since some time that effects of new weakly coupled physics at a high scale $\Lambda$ can be described by an effective Lagrangian~\cite{Burges:1983zg,Leung:1984ni,Buchmuller:1985jz}
\begin{equation}
\mathcal{L}^\text{eff} = \sum \frac{C_x}{\Lambda^2} O_x + \dots \,,
\label{ec:effL}
\end{equation}
where $O_x$ are dimension-six gauge-invariant operators and $C_x$ are complex constants. The dots in the above equation stand for higher-dimension operators neglected in this work, whose contributions are suppressed by higher powers of $\Lambda$. The framework of effective operators is very convenient to describe the effect of new physics in fermion couplings. In particular, for the top quark the effective operator contributions have been already obtained, for third generation
couplings~\cite{Gounaris:1996vn,Gounaris:1996yp,Whisnant:1997qu,Yang:1997iv} and also including intergenerational mixing~\cite{delAguila:2000aa,delAguila:2000rc,Ferreira:2008cj,
Coimbra:2008qp,Ferreira:2009bf,AguilarSaavedra:2008zc}.

The usefulness of the effective operator approach arises from the fact that any new physics contribution (in our case, at order $1/\Lambda^2$) can be parameterised in terms of a complete operator basis, and this framework is then completely general. But it is also very important to have such a basis in a minimal form.
In this direction, a great effort was done in Ref.~\cite{Buchmuller:1985jz} to remove redundant operators and reduce the set of independent $O_x$. (This can always be done by using the equations of motion, even for off-shell particles or when loop corrections are considered~\cite{Georgi:1991ch,Arzt:1993gz}.)
Still, few redundant ones remained, as pointed out in Ref.~\cite{Grzadkowski:2003tf}, and recently the set of independent operators has been further reduced \cite{AguilarSaavedra:2008zc}. This latter simplification has been particularly useful, because it implies that the most general fermion-fermion-gauge boson interactions,
including the SM contributions as well as those from dimension-six operators, can be parameterised in full generality with only $\gm$ and $\smn q_\nu$ terms, where $q_\nu$ is the gauge boson momentum. In this paper we extend previous work and obtain new operator equalities which allow to simplify the general form of the $H f_i f_j$ interactions among two fermions $f_i$, $f_j$ and the Higgs boson. For brevity, we concentrate ourselves on $Htt$ and $Htq$ interactions, although the results are general for all fermions. Among the operators listed
in Ref.~\cite{Buchmuller:1985jz}, those contributing to the $Htt$ and $Htq$ vertices are
\begin{eqnarray}
O_{u\phi}^{ij} & = & (\phi^\dagger \phi) (\bar q_{Li} u_{Rj} \tilde \phi ) \,, \notag \\
O_{\phi q}^{(3,ij)} & = & i (\phi^\dagger \tau^I D_\mu  \phi)
       (\bar q_{Li} \gm \tau^I q_{Lj}) \,, \notag \\
O_{\phi q}^{(1,ij)} & = & i (\phi^\dagger D_\mu \phi) (\bar q_{Li} \gm q_{Lj}) \,, \notag \\
O_{\phi u}^{ij} & = & i (\phi^\dagger D_\mu \phi) (\bar u_{Ri} \gm u_{Rj}) \,,
\label{ec:Oall}
\end{eqnarray}
where we have already omitted several operators already shown to be redundant \cite{AguilarSaavedra:2008zc}. (We denote the quark fields as $q_{Li}$, $u_{Ri}$, $d_{Ri}$ in standard notation, with $i,j=1,2,3$ flavour indices.) The first operator in Eqs.~(\ref{ec:Oall}) gives scalar $\bar f_i f_j H$ and pseudo-scalar $\bar f_i \gamma_5 f_j H$ contributions to the vertices, while the three remaining ones give derivative terms of the type $\bar f_i \gm P_{L,R} f_j \pam H$. However, in this paper we will show that these derivative interactions are redundant and can be dropped. Indeed, instead of using the operators
\begin{equation}
O_{\phi q}^{(3,ij)} \,,\quad O_{\phi q}^{(1,ij)} \,,\quad O_{\phi u}^{ij}  \,,
\label{ec:Oij}
\end{equation}
with $i,j=1,2,3$
to parameterise new physics contributions, one may equivalently use the sums and differences
of operators with $i \leq j = 1,2,3$,
\begin{eqnarray}
O_{\phi q}^{(3,i+j)} & \equiv & \frac{1}{2} \left[ O_{\phi q}^{(3,ij)} + 
(O_{\phi q}^{(3,ji)})^\dagger \right] \,, \notag \\
O_{\phi q}^{(1,i+j)} & \equiv & \frac{1}{2} \left[ O_{\phi q}^{(1,ij)} + 
(O_{\phi q}^{(1,ji)})^\dagger \right] \,, \notag \\
O_{\phi u}^{i+j} & \equiv & \frac{1}{2} \left[ O_{\phi u}^{ij} +
(O_{\phi u}^{ji})^\dagger \right] \,, \notag \\
O_{\phi q}^{(3,i-j)} & \equiv & \frac{1}{2} \left[ O_{\phi q}^{(3,ij)} - 
(O_{\phi q}^{(3,ji)})^\dagger \right] \,, \notag \\
O_{\phi q}^{(1,i-j)} & \equiv & \frac{1}{2} \left[ O_{\phi q}^{(1,ij)} - 
(O_{\phi q}^{(1,ji)})^\dagger \right] \,, \notag \\
O_{\phi u}^{i-j} & \equiv & \frac{1}{2} \left[ O_{\phi u}^{ij} -
(O_{\phi u}^{ji})^\dagger \right] \,.
\label{ec:Onew}
\end{eqnarray}
For $i \neq j$ the equivalence of both sets is evident, whereas for $i=j$ the change of basis amounts to decomposing the operators in Eq.~(\ref{ec:Oij}) (with complex coefficients in general) into their hermitian and anti-hermitian parts (with real and purely imaginary coefficients, respectively). As we will show, the opposite-sign combinations in Eqs.~(\ref{ec:Onew}) are redundant and therefore they can be dropped from the operator list. For a pair of different flavour indices $i \neq j$ this reduces the number of relevant operators: the six operators
$O_x^{ij}$, $O_x^{ji}$ in Eq.~(\ref{ec:Oij}) are replaced by the first three ones in Eqs.~(\ref{ec:Onew}), namely $O_{x}^{i+j}$ with $i < j$. For the same flavour $i=j$ the three operators $O_x^{ii}$
are replaced by their hermitian parts $O_x^{i+i}$. The advantage of this change of basis, apart from the reduction in the number of relevant operators, is that the same-sign
combinations $O_x^{i+j}$ in Eqs.~(\ref{ec:Onew}) do not give $H f_i f_j$ vertices 
because the derivative terms originating from $O_x^{ij}$ and $(O_x^{ji})^\dagger$ exactly cancel. Hence, the structure of these vertices is reduced to scalar and pseudo-scalar terms.

It is important to remark here the importance of simplifying the general form of the fermion interactions with the gauge and Higgs bosons. First, because the number of relevant parameters necessary to describe the vertices is reduced, with an obvious advantage for the study of anomalous interactions and the extraction of limits from experimental data: the dimensionality of the parameter space is reduced and the analyses are greatly simplified. Additionally, the presence of more variables than the ones really needed would lead to ``perfect'' correlations in the limits obtained on them, which apparently would suggest that certain combinations of variables cannot be experimentally measured. However, these correlations only reflect the fact that some parameters can actually be eliminated from the analysis. A second advantage concerns Monte Carlo generation of the new signals, because with an adequate choice of gauge-invariant operators the number of diagrams and anomalous interactions contributing to a given process is reduced, which is quite useful for the development of generators for non-SM processes. In the case of the top quark, generators for top production processes involving anomalous couplings~\cite{AguilarSaavedra:2008gt} benefit from the simplification of the interactions with the $W$ and $Z$ bosons, the photon and the gluon.
One must note, however, that different operator bases may always be choosen to parameterise new physics effects~\cite{Ferreira:2008cj,Coimbra:2008qp,Ferreira:2009bf}, and results for cross sections, branching ratios, etc. will clearly be basis-independent. But it is also clear that an adequate selection minimising the number of parameters necessary to describe the most relevant top quark production and decay processes considerably simplifies the calculations.

In this paper we first prove in section~\ref{sec:2} the operator equalities which allow to drop the opposite-sign combinations in Eqs.~(\ref{ec:Onew}). Then, we present in section~\ref{sec:3} the minimal $Htt$ and $Htq$ vertices, and in section~\ref{sec:4} we discuss how the operator replacements made affect gauge boson vertices. Although the structure of the latter is not simplified with respect to previous work~\cite{AguilarSaavedra:2008zc}, the number of contributing operators decreases in some cases. In section~\ref{sec:5} we summarise our results. Appendix~\ref{sec:a} studies in detail quark mixing. Since the equalities obtained in section~\ref{sec:2} relate operators which do not contribute to quark masses to other ones which do, understanding their implications for $Htt$, $Htq$ vertices requires a correct account of quark mixing effects. In appendix~\ref{sec:b} it is shown with an explicit example that the trilinear vertex replacements dictated by the operator equalities give the same results in amplitude calculations. Finally, the contributions of the relevant effective operators to the $Htt$ and $Htq$ vertices, as well as to quark masses, are collected
in appendix~\ref{sec:c}.

\section{Effective operator equalities}
\label{sec:2}

In this paper we approximately follow the notation in Ref.~\cite{Buchmuller:1985jz}. The conventions and signs used for covariant derivatives, the equations of motion, etc. are given in detail in Ref.~\cite{AguilarSaavedra:2008zc}.
Using the definitions in Eqs.~(\ref{ec:Oall}), the last three operators in Eqs.~(\ref{ec:Onew}) can be written as
\begin{eqnarray}
O_{\phi q}^{(3,i-j)} & = & \frac{i}{2} \left[ \phi^\dagger \tau^I D_\mu \phi
+ (D_\mu \phi)^\dagger \tau^I \phi \right] (\bar q_{Li} \gm \tau^I q_{Lj}) \,, \notag \\
O_{\phi q}^{(1,i-j)} & = & \frac{i}{2} \left[ \phi^\dagger D_\mu \phi
+ (D_\mu \phi)^\dagger \phi \right] (\bar q_{Li} \gm q_{Lj})
= \frac{i}{2} D_\mu (\phi^\dagger \phi) (\bar q_{Li} \gm q_{Lj}) \,, \notag \\
O_{\phi u}^{i-j} & = & \frac{i}{2} \left[ \phi^\dagger D_\mu \phi
+ (D_\mu \phi)^\dagger \phi \right] (\bar u_{Ri} \gm u_{Rj})
= \frac{i}{2} D_\mu (\phi^\dagger \phi) (\bar u_{Ri} \gm u_{Rj}) \,.
\end{eqnarray}
Integrating by parts, and using for the first operator the equality
\begin{equation}
D_\mu \tau^I \psi = \tau^I D_\mu \psi + g \epsilon_{IJK} \Wm^J \tau^K \psi \,,
\end{equation}
for $\psi=\phi,q_{Lj}$ (or any $\text{SU}(2)$ doublet), we have
\begin{eqnarray}
O_{\phi q}^{(3,i-j)} & = & \frac{1}{2} \left[ (\phi^\dagger \tau^I \phi) \overline{(i \Dsl q_{Li})} \tau^I q_{Lj} - (\phi^\dagger \tau^I \phi) \bar q_{Li} \tau^I (i \Dsl q_{Lj}) \right] \,, \notag \\
O_{\phi q}^{(1,i-j)} & = & \frac{1}{2} \left[ (\phi^\dagger \phi) \overline{(i \Dsl q_{Li})} q_{Lj} - (\phi^\dagger \phi) \bar q_{Li} (i \Dsl q_{Lj}) \right] \,, \notag \\
O_{\phi u}^{i-j} & = & \frac{1}{2} \left[ (\phi^\dagger \phi) \overline{(i \Dsl u_{Ri})} u_{Rj} 
  - (\phi^\dagger \phi) \bar u_{Ri} (i \Dsl u_{Rj}) \right] \,.
\end{eqnarray}
Then, using the quark equations of motion it is then easy to find that
\begin{eqnarray}
O_{\phi q}^{(3,i-j)} & = &  \frac{1}{2} \left[ Y_{jk}^u O_{u\phi}^{ik} - Y_{jk}^d O_{d\phi}^{ik}
- Y_{ki}^{u\dagger} (O_{u\phi}^{jk})^\dagger + Y_{ki}^{d\dagger} (O_{d\phi}^{jk})^\dagger
\right] \,, \notag \\
O_{\phi q}^{(1,i-j)} & = & \frac{1}{2} \left[ - Y_{jk}^u O_{u\phi}^{ik} - Y_{jk}^d O_{d\phi}^{ik}
+ Y_{ki}^{u\dagger} (O_{u\phi}^{jk})^\dagger + Y_{ki}^{d\dagger} (O_{d\phi}^{jk})^\dagger
\right] \,, \notag \\
O_{\phi u}^{i-j} & = & \frac{1}{2} \left[ Y_{ki}^u O_{u\phi}^{kj} 
- Y_{jk}^{u\dagger} (O_{u\phi}^{ki})^\dagger \right] \,,
\label{ec:Oeq}
\end{eqnarray}
where $Y^u$, $Y^d$ are the up and down quark Yukawa matrices, respectively, and
\begin{equation}
O_{d\phi}^{ij} = (\phi^\dagger \phi) (\bar q_{Li} d_{Rj} \phi ) \,.
\end{equation}
These relations show that the operators on the left-hand side are redundant.
Notice that for $O_{\phi q}^{(3,i-j)}$ the terms involving Pauli matrices can be simplified after using the equations of motion, because
\begin{equation}
(\psi_1^\dagger \tau^I \psi_2) (\psi_3^\dagger \tau^I \psi_2) = (\psi_1^\dagger \psi_2) (\psi_3^\dagger \psi_2) \,,
\end{equation}
being $\psi_i$ any $\text{SU}(2)$ doublets.

The operator equalities in Eqs.~(\ref{ec:Oeq}) imply in particular that, in the calculation of gauge-invariant quantities, the derivative contribution to the top-Higgs trilinear vertices from the operators $O_{\phi q}^{(3,i-j)}$, $O_{\phi q}^{(1,i-j)}$, $O_{\phi u}^{i-j}$ can be replaced by the scalar and pseudo-scalar contributions from the right-hand side of the corresponding equations, provided that all the remaining operator contributions, if any, are included. It must be noticed, however, that Eqs.~(\ref{ec:Oeq})
relate some operators (on the left-hand side) which do not contribute to quark masses to some others which do contribute (the combinations on the right-hand side). Therefore, quark mixing plays a central role in the correct extraction of relations among trilinear vertices from the gauge-invariant operator equalities. This issue is discussed in detail in appendix~\ref{sec:a}.
In appendix~\ref{sec:b} it is shown with an explicit example that the derivative terms can 
indeed be replaced by (pseudo-)scalar terms in amplitude calculations.

Finally, it is worth pointing out that for brevity we have restricted ourselves to operators involving the top quark, but for other operators involving only down-type quarks or leptons,
\begin{align}
& O_{\phi d}^{ij} = i (\phi^\dagger D_\mu \phi) (\bar d_{Ri} \gm d_{Rj}) \,, \notag \\
& O_{\phi \ell}^{(3,ij)} = i (\phi^\dagger \tau^I D_\mu  \phi)
       (\bar \ell_{Li} \gm \tau^I \ell_{Lj}) \,, \notag \\
& O_{\phi \ell}^{(1,ij)} = i (\phi^\dagger D_\mu \phi) (\bar \ell_{Li} \gm \ell_{Lj}) \,, \notag \\
& O_{\phi e}^{ij} = i (\phi^\dagger D_\mu \phi) (\bar e_{Ri} \gm e_{Rj}) \,,
\label{ec:Onot}
\end{align}
the same reckoning applies, and only the sums $O_x^{ij} + (O_x^{ji})^\dagger$ are relevant.

\section{General $Htt$ and $Htq$ interactions}
\label{sec:3}

We give here general expressions for the $Htt$ and $Htq$ vertices arising from dimension-six gauge-invariant operators, excluding the redundant ones. The Lagrangians given here are valid
for off-shell particles and do not make any assumption about fermion masses and mixings, because the proof that $O_{\phi q}^{(3,i-j)}$, $O_{\phi q}^{(1,i-j)}$
and $O_{\phi u}^{i-j}$ are redundant involves arbitrary Yukawa matrices.
The contributions to the $Htt$ and $Htq$ vertices of the effective operators in Eqs.~(\ref{ec:Oall}) are collected in appendix \ref{sec:c} for complex coefficients $C_x$.

\subsection{$Htt$ vertex}

For $i=j=3$, one may in full generality replace the gauge-invariant operators $O_{\phi q}^{(3,33)}$, $O_{\phi q}^{(1,33)}$ and $O_{\phi u}^{33}$ by the hermitian ones
\begin{eqnarray}
O_{\phi q}^{(3,3+3)} & = & \frac{i}{2} \left[ \phi^\dagger \tau^I D_\mu \phi
- (D_\mu \phi)^\dagger \tau^I \phi \right] (\bar q_{L3} \gm \tau^I q_{L3}) \,, \notag \\
O_{\phi q}^{(1,3+3)} & = & \frac{i}{2} ( \phi^\dagger \DMD \phi) (\bar q_{L3} \gm q_{L3}) \,, \notag \\
O_{\phi u}^{3+3} & = &  \frac{i}{2} ( \phi^\dagger \DMD \phi) (\bar u_{R3} \gm u_{R3}) \,.
\label{ec:Oplus33}
\end{eqnarray}
These substitutions simplify the form of the $Htt$ interaction, because the contributions of the new operators to this vertex identically vanish. Thus, the most general $Htt$ vertex including the SM Yukawa coupling as well as corrections from dimension-six operators is 
\begin{equation}
\mathcal{L}_{Htt} = - \frac{1}{\sqrt 2} \, \bar t \left( \yttv +
 i \ytta \gamma_5  \right) t\; H \,.
\label{ec:Htt}
\end{equation}
with $\yttv = \sqrt 2 m_t/v$, $\ytta = 0$ in the SM at the tree level. The contributions from dimension-six operators are
\begin{eqnarray}
\delta \yttv & = & -\frac{3}{2} \, \RE C_{u\phi}^{33} \frac{v^2}{\Lambda^2}  \,, \nonumber \\
\delta \ytta & = & -\frac{3}{2} \, \IM C_{u\phi}^{33} \frac{v^2}{\Lambda^2} \,.
\end{eqnarray}
We point out that this effective Lagrangian, while being completely general, is far simpler than the one given in previous literature~\cite{Whisnant:1997qu,Yang:1997iv} which involves several derivative terms.

\subsection{$Htq$ vertices}

For different flavour indices $i \neq j$, the six operators
$O_{\phi q}^{(3,ij)}$, $O_{\phi q}^{(1,ij)}$ and $O_{\phi u}^{ij}$ can be replaced by the three same-sign combinations
\begin{eqnarray}
O_{\phi q}^{(3,i+j)} & = & \frac{i}{2} \left[ \phi^\dagger \tau^I D_\mu \phi
- (D_\mu \phi)^\dagger \tau^I \phi \right] (\bar q_{Li} \gm \tau^I q_{Lj}) \,, \notag \\
O_{\phi q}^{(1,i+j)} & = & \frac{i}{2} ( \phi^\dagger \DMD \phi) (\bar q_{Li} \gm q_{Lj}) \,, \notag \\
O_{\phi u}^{i+j} & = &  \frac{i}{2} ( \phi^\dagger \DMD \phi) (\bar u_{Ri} \gm u_{Rj}) \,.
\label{ec:Oplusij}
\end{eqnarray}
This reduces the number of operators relevant for FCN interactions and, more importantly, it simplifies the form of the $Htq$ vertices because the operators in Eqs.~(\ref{ec:Oplusij}) do not contribute (the contributions of both terms identically cancel). 
Then, the most general flavour-changing $Htc$ vertex arising from dimension-six effective operators can be written as
\begin{equation}
\mathcal{L}_{Htc} = - \frac{1}{\sqrt 2} \, \bar c \left( \yctl P_L +
  \yctr P_R  \right) t\; H + \text{H.c.} \,,
\label{ec:Htc}
\end{equation}
where $\yctl = \yctr = 0$ in the SM at the tree level. The contributions from dimension-six operators are
\begin{eqnarray}
\delta \yctl & = & -\frac{3}{2} \, C_{u\phi}^{32*} \frac{v^2}{\Lambda^2}  \,, \notag \\
\delta \yctr & = & -\frac{3}{2} \, C_{u\phi}^{23} \frac{v^2}{\Lambda^2} \,.
\end{eqnarray}
For the $Htu$ vertex the results are analogous by changing the flavour indices $2 \to 1$. These effective Lagrangians are also simpler than the ones previously found~\cite{delAguila:2000rc,delAguila:2000aa}. We remark here that equivalent Lagrangians have been used to study top FCN processes induced by $Htq$ couplings~\cite{AguilarSaavedra:2000aj}, namely top pair production with FCN decay of one top quark and associated $Ht$ production. Those studies are thus completely general from the point of view of dimension-six effective operators.

\section{Contributions to gauge boson vertices}
\label{sec:4}

The change of operator basis performed with the substitution of $O_{\phi q}^{(3,ij)}$, $O_{\phi q}^{(1,ij)}$ and $O_{\phi u}^{ij}$ by
$O_{\phi q}^{(3,i+j)}$, $O_{\phi q}^{(1,i+j)}$ and $O_{\phi u}^{i+j}$
does not modify the general Lorentz structure of the top quark gauge interactions, which only involve $\gm$ and $\smn q_\nu$ couplings as shown in Ref.~\cite{AguilarSaavedra:2008zc}.
For photon and gluon interactions the effective Lagrangians remain the same because
$O_{\phi q}^{(3,ij)}$, $O_{\phi q}^{(1,ij)}$ and $O_{\phi u}^{ij}$ do not contribute. For the $W$ and $Z$ boson interactions there are few modifications. The
additional operators contributing to these vertices but not to Higgs couplings are
\begin{eqnarray}
O_{\phi \phi}^{ij} & = & i (\tilde \phi^\dagger D_\mu \phi) (\bar u_{Ri} \gm d_{Rj}) \,, \notag \\
O_{uW}^{ij} & = & (\bar q_{Li} \smn \tau^I u_{Rj}) \tilde \phi \, \Wmn^I \,, \notag \\
O_{dW}^{ij} & = & (\bar q_{Li} \smn \tau^I d_{Rj}) \phi \, \Wmn^I \,, \notag \\
O_{uB\phi}^{ij} & = & (\bar q_{Li} \smn u_{Rj}) \tilde \phi \, \Bmn \,.
\end{eqnarray}
The $Wtb$ vertex is still parameterised as~\cite{AguilarSaavedra:2006fy}
\begin{eqnarray}
\mathcal{L}_{Wtb} & = & - \frac{g}{\sqrt 2} \bar b \, \gamma^{\mu} \left( \vl P_L
  + \vr P_R \right) t\; W_\mu^- 
   \nonumber \\
& & - \frac{g}{\sqrt 2} \bar b \, \frac{i \sigma^{\mu \nu} q_\nu}{M_W}
  \left( \gl P_L + \gr P_R \right) t\; W_\mu^- + \mathrm{H.c.}
\label{ec:Wtb}
\end{eqnarray}
Within the SM, $\vl$ equals the Cabibbo-Kobayaski-Maskawa matrix element $V_{tb}\simeq 1$, while the rest of couplings
$\vr$, $\gl$ and $\gr$ vanish at the tree level. The contributions to these couplings from dimension-six gauge-invariant operators are
\begin{align}
& \delta \vl = C_{\phi q}^{(3,3+3)*}
     \frac{v^2}{\Lambda^2}   \,, 
&& \delta \gl = \sqrt 2 C_{dW}^{33*} \frac{v^2}{\Lambda^2} 
  \,, \notag \\
&  \delta \vr = \frac{1}{2} C_{\phi \phi}^{33} \frac{v^2}{\Lambda^2} \,,
&& \delta \gr = \sqrt 2 C_{uW}^{33} \frac{v^2}{\Lambda^2}  \,.
\end{align}
The only difference with respect to Ref.~\cite{AguilarSaavedra:2008zc} is the replacement of the operator coefficient
\begin{equation}
C_{\phi q}^{(3,33)} \to C_{\phi q}^{(3,3+3)} \,.
\label{ec:Csub1}
\end{equation}
Notice however that since $O_{\phi q}^{(3,3+3)}$ is hermitian, its coefficient
$C_{\phi q}^{(3,3+3)}$ is real in order to ensure the hermiticity of the Lagrangian.
The other contributions to the $Wtb$ effective vertex are complex in general.
The $Ztt$ vertex is parameterised as \cite{AguilarSaavedra:2008zc}
\begin{eqnarray}
\mathcal{L}_{Ztt} & = & - \frac{g}{2 c_W} \bar t \, \gm \left( \xttl P_L
  + \xttr P_R - 2 s_W^2 Q_t \right) t\; Z_\mu  \nonumber \\
& & - \frac{g}{2 c_W} \bar t \, \frac{i \smn q_\nu}{M_Z}
  \left( \dvz + i \daz \gamma_5 \right) t\; Z_\mu  \,,
\label{ec:Ztt}
\end{eqnarray}
with $Q_t=2/3$ the top quark electric charge.
Within the SM, these couplings take the values $\xttl = 2 \, T_3(t_L) = 1$, $\xttr = 2 \, T_3 (t_R) = 0$, where $T_3$ denotes the third isospin component, and $\dvz = \daz = 0$ at the tree level. The contributions from dimension-six operators are
\begin{align}
& \delta \xttl = \left[  C_{\phi q}^{(3,3+3)} - C_{\phi q}^{(1,3+3)}
\right]  \frac{v^2}{\Lambda^2} \,, \notag \\
& \delta \xttr = - C_{\phi u}^{3+3} \frac{v^2}{\Lambda^2} \,, \notag \\
& \delta \dvz = \sqrt 2 \, \RE \left[ c_W  C_{uW}^{33} - s_W C_{uB\phi}^{33} \right]  \frac{v^2}{\Lambda^2} \,, \notag \\
& \delta \daz = \sqrt 2 \, \IM \left[ c_W C_{uW}^{33} - s_W  C_{uB\phi}^{33} \right] \frac{v^2}{\Lambda^2} \,.
\end{align}
The differences with respect to the results in Ref.~\cite{AguilarSaavedra:2008zc} are the replacements
\begin{eqnarray}
\RE \, C_{\phi q}^{(3,33)} & \to & C_{\phi q}^{(3,3+3)} \,, \notag \\
\RE \, C_{\phi q}^{(1,33)} & \to & C_{\phi q}^{(1,3+3)} \,, \notag \\
\RE \, C_{\phi u}^{33} & \to & C_{\phi u}^{3+3} \,,
\label{ec:Csub2}
\end{eqnarray}
which do not reduce the number of operators involved nor the number of parameters needed to describe the $Ztt$ vertex, because the coefficients $C_x$ on the right column of Eqs.~(\ref{ec:Csub2}) are real.
Finally, for $Ztq$ vertices the number of relevant contributing operators is reduced.
The $Ztc$ vertex can still be parameterised with the Lagrangian~\cite{AguilarSaavedra:2008zc}
\begin{eqnarray}
\mathcal{L}_{Ztc} & = & - \frac{g}{2 c_W} \bar c \, \gm \left( \xctl P_L
  + \xctr P_R \right) t\; Z_\mu  \nonumber \\
& & - \frac{g}{2 c_W} \bar c \, \frac{i \smn q_\nu}{M_Z}
  \left( \kl P_L + \kr P_R \right) t\; Z_\mu  + \text{H.c.}  \,,
\label{ec:Ztc}
\end{eqnarray}
but the anomalous couplings depend on a smaller number of effective operator coefficients (five instead of eight),
\begin{align}
& \delta \xctl = \frac{1}{2} \left[ C_{\phi q}^{(3,2+3)} 
      - C_{\phi q}^{(1,2+3)} \right] \frac{v^2}{\Lambda^2}
\,, \notag \\
& \delta \xctr = - \frac{1}{2} C_{\phi u}^{2+3} \frac{v^2}{\Lambda^2}
\,, \notag \\
& \delta \kl = \sqrt 2 \left[ c_W C_{uW}^{32*} - s_W C_{uB\phi}^{32*} 
  \right] \frac{v^2}{\Lambda^2}
\,, \notag \\
& \delta \kr = \sqrt 2 \left[ c_W C_{uW}^{23} - s_W C_{uB\phi}^{23} \right]
 \frac{v^2}{\Lambda^2}
\,,
\label{ec:Ztcex}
\end{align}
We see that the differences with respect to the results in Ref.~\cite{AguilarSaavedra:2008zc} are the replacements
\begin{eqnarray}
C_{\phi q}^{(3,23)} + C_{\phi q}^{(3,32)*} & \to & C_{\phi q}^{(3,2+3)} \,, \notag \\
C_{\phi q}^{(1,23)} + C_{\phi q}^{(1,32)*} & \to & C_{\phi q}^{(1,2+3)} \,, \notag \\
C_{\phi u}^{23} + C_{\phi u}^{32*} & \to & C_{\phi u}^{2+3} \,.
\end{eqnarray}
For $Ztu$ couplings the Lagrangian is similar, and the effective operator contributions are obtained by replacing the flavour indices $2 \to 1$.

\section{Summary}
\label{sec:5}

If the Higgs boson is found at LHC, the experimental measurement of its properties and couplings, in particular those involving the top quark, will be of utmost importance to investigate whether it is the SM Higgs or there is any new physics associated to electroweak symmetry breaking. Hence, an adequate parameterisation of the most general Higgs interactions with fermions is fundamental, and finding a minimal one involving the smallest parameter set is very useful. With such a minimal set the interpretation of experimental measurements in terms of the underlying physics, {\em e.g.} limits on new physics at a higher scale, is considerably simplified.

In this paper we have shown that the fermion-fermion-Higgs interactions arising from dimension-six gauge-invariant effective operators can be described in full generality using only scalar 
$\bar f_i f_j H$ and pseudo-scalar $\bar f_i \gamma_5 f_j H$ terms. The key for this useful result is the derivation of new relations among gauge-invariant operators using the equations of motion. These relations allow to drop several redundant operators which give 
derivative terms of the type $\bar f_i \gm P_{L,R} f_j \pam H$, while the remaining operators do not give such contributions. We have given explicit examples for the top quark, namely the $Htt$, $Htu$ and $Htc$ vertices, writing the anomalous couplings in terms of coefficients of gauge-invariant operators.

The new operator relations found do not simplify the structure of the fermion-fermion-gauge boson vertices which, as previously shown \cite{AguilarSaavedra:2008zc}, involve only $\gm$ and $\smn q_\nu$ couplings, with $q_\nu$ the boson momentum. Nevertheless, for the $Wtb$ vertex they imply that the correction from effective operators to the SM coupling $V_{tb}$ is real, and
for $Ztu$, $Ztc$ interactions the number of relevant effective operators contributing to each effective vertex is reduced from eight to five. 

The simplification of the top-Higgs vertices, which completes the study made for top gauge interactions, is also very useful to build Monte Carlo generators for new processes beyond the SM. In this way, single top production through FCN couplings and top pair production with
FCN decay of one top quark have been added to the {\tt Protos} generator~\cite{AguilarSaavedra:2008gt}, using the minimal sets of anomalous $Ztq$, $\gamma tq$, $gtq$ and $Htq$ couplings obtained. The advantage of these minimal sets is that they greatly simplify the calculations for the most relevant top production processes at LHC, reducing the number of parameters, Lorentz structures and diagrams required in a gauge-invariant calculation of amplitudes.
But, more importantly, the dimensionality of the parameter space to be investigated (and on which constraints are to be placed) is reduced roughly by a factor of two, which is a tremendous advantage for phenomenological studies.

\section*{Acknowledgements}

I thank M. P\'erez-Victoria for useful comments and for reading the manuscript.
This work has been supported by a MEC Ram\'on y Cajal contract, MEC project FPA2006-05294 and
Junta de Andaluc{\'\i}a projects FQM 101 and FQM 437.

\appendix

\section{Operator equalities and quark mixing}
\label{sec:a}

Once that it has been shown that the operators $O_{\phi q}^{(3,i-j)}$, $O_{\phi q}^{(1,i-j)}$ and $O_{\phi u}^{i-j}$ are redundant, they can be removed from the operator basis in the same way as many other redundant operators are~\cite{Buchmuller:1985jz}. Still, it is illustrating to see how the operator equalities in Eqs.~(\ref{ec:Oeq}) translate into relations among $H f_i f_j$ couplings. In other words, we are interested in investigating how the derivative terms included in these operators can be replaced by (pseudo-)scalar terms in the calculation of gauge-invariant quantities.
Quark mixing plays a central role in this derivation: while $O_{\phi q}^{(3,i-j)}$, $O_{\phi q}^{(1,i-j)}$ and $O_{\phi u}^{i-j}$ do not contribute to quark masses, the combinations obtained from them using the equations of motion give non-diagonal corrections to the mass matrices.
In this appendix we obtain such replacement relations for the simplest case, the operator $O_{\phi u}^{i-j}$, and taking $i=2$, $j=3$ for definiteness. The remaining cases are completely analogous.
In appendix~\ref{sec:b} we show with an example that such replacements give the same result in amplitude calculations.

We can assume without loss of generality that the up-quark Yukawa matrix is diagonal. (In absence of dimension-six operator contributions to quark masses, this implies that the weak and mass eigenstates coincide.) Then, the operator equality for $O_{\phi u}^{2-3}$ is
\begin{equation}
O_{\phi u}^{2-3} \equiv \frac{1}{2} \left[ O_{\phi u}^{23} - (O_{\phi u}^{32})^\dagger 
\right] = \frac{1}{2} \left[ Y_{22}^u O_{u\phi}^{23} - Y_{33}^{u*} (O_{u\phi}^{32})^\dagger \right] \,.
\label{ec:Oeq3}
\end{equation}
For brevity, we denote as $\bar O_{\phi u}^{2-3}$ the right-hand side of this equation,
\begin{equation}
\bar O_{\phi u}^{2-3} \equiv \frac{1}{2} \left[ Y_{22}^u O_{u\phi}^{23} - Y_{33}^{u*} (O_{u\phi}^{32})^\dagger \right] \,.
\end{equation}
The operator $O_{\phi u}^{2-3}$ does not give corrections to quark masses. Its contribution to the $Htc$ vertex is (see appendix~\ref{sec:c})
\begin{equation}
\alpha O_{\phi u}^{2-3} + \alpha^* (O_{\phi u}^{2-3})^\dagger \supset i \frac{v}{2}
\left[ \alpha \, \bar c_R \gm t_R - \alpha^* \, \bar t_R \gm c_R \right] \pam H \,,
\label{ec:cont1}
\end{equation}
where $\alpha$ stands for $C_{\phi u}^{2-3}/\Lambda^2$. On the other hand, $\bar O_{\phi u}^{2-3}$ induces non-diagonal Yukawa terms,
\begin{equation}
\alpha \bar O_{\phi u}^{2-3} + \alpha^* (\bar O_{\phi u}^{2-3})^\dagger \supset
\frac{v^3}{4 \sqrt 2} \left[ \alpha (Y_{22}^u \, \bar c'_L t'_R - Y_{33}^{u*} \bar c'_R t'_L) +
\alpha^* (Y_{22}^{u*} \bar t'_R c'_L - Y_{33}^u \, \bar t'_L c'_R)
\right] \,.
\end{equation}
In this case we denote the weak eigenstates with primes to distinguish them from the mass eigenstates, which are obtained after diagonalising the mass matrix for up-type quarks. Introducing the small dimensionless parameter $\varepsilon = \alpha v^2/4 = C_{\phi u}^{2-3} (v/\Lambda)^2 /4$, the mass term for up-type quarks is
\begin{equation}
\mathcal{L}_\text{mass} = - ( \bar u'_L \; \bar c'_L \; \bar t'_L ) \frac{v}{\sqrt 2}
\left( \! \begin{array}{ccc} Y_{11}^u & 0 & 0 \\
0 & Y_{22}^u & \varepsilon Y_{22}^u \\ 0 & -\varepsilon^* Y_{33} & Y_{33}
\end{array} \! \right)
\left( \! \begin{array}{c} u'_R \\ c'_R \\ t'_R \end{array} \! \right) + \text{H.c.}
\end{equation}
At first order in $\varepsilon$, the relation between weak and mass eigenstates is
\begin{equation}
u'_{L,R} = u_{L,R} \,;~ c'_L = c_L \,,~ c'_R = c_R - \varepsilon t_R \,;~ t'_L=t_L \,,~ 
 t'_R = t_R + \varepsilon^* c_R \,. 
\end{equation}
We can now compute the contribution of $\bar O_{\phi u}^{2-3}$ to the $Htc$ vertex. The first one comes from the operator itself,
\begin{equation}
\alpha \bar O_{\phi u}^{2-3} + \alpha^* (\bar O_{\phi u}^{2-3})^\dagger \supset
\frac{3 v^2}{4 \sqrt 2} \left[ \alpha (Y_{22}^u \, \bar c_L t_R - Y_{33}^{u*} \bar c_R t_L) +
\alpha^* (Y_{22}^{u*} \bar t_R c_L - Y_{33}^u \, \bar t_L c_R)
\right] H \,.
\label{ec:cont2a}
\end{equation}
Notice that in this equation we have replaced weak by mass eigenstates at first order in $\varepsilon$, because they are already multiplied by terms of order $\varepsilon$. The second contribution to the $Htc$ vertex comes after writing the
SM Yukawa interactions, which are diagonal in the primed basis,
\begin{eqnarray}
\mathcal{L}_\text{Y} & = & - \frac{1}{\sqrt 2} \left[ Y_{11}^u \, \bar u'_L u'_R + Y_{22}^u \, \bar c'_L c'_R + Y_{33}^u \, \bar t'_L t'_R \right] H \notag \\
& & - \frac{1}{\sqrt 2} \left[ Y_{11}^{u*} \, \bar u'_R u'_L + Y_{22}^{u*} \, \bar c'_R c'_L + Y_{33}^{u*} \, \bar t'_R t'_L \right] H
\end{eqnarray}
in terms of the mass eigenstates $u_{L,R}$, $c_{L,R}$, $t_{L,R}$. This additional contribution coming from mixing is
\begin{equation}
\frac{v^2}{4 \sqrt 2} \left[
\alpha (- Y_{22}^u \bar c_L t_R + Y_{33}^{u*} \bar c_R t_L )
+ \alpha^* (-Y_{22}^{u*} \bar t_R c_L + Y_{33}^u \bar t_L c_R) \right] H \,,
\label{ec:cont2b}
\end{equation}
so that the total contribution when $\alpha \bar O_{\phi u}^{2-3} + \alpha^* (\bar O_{\phi u}^{2-3})^\dagger$ is added to the SM Lagrangian is
\begin{equation}
\frac{v^2}{2 \sqrt 2} \left[ \alpha (Y_{22}^u \, \bar c_L t_R - Y_{33}^{u*} \bar c_R t_L) +
\alpha^* (Y_{22}^{u*} \bar t_R c_L - Y_{33}^u \, \bar t_L c_R)
\right] H \,.
\label{ec:cont2}
\end{equation}
Comparing Eqs.~(\ref{ec:cont1}) and (\ref{ec:cont2}),
and introducing the quark masses $m_c=Y_{22}^u v/\sqrt 2$, $m_t=Y_{33}^u v/\sqrt 2$,
we observe that the operator equality implies the replacements
\begin{eqnarray}
i \bar c_R \gm t_R \pam H & \to & (m_c \bar c_L t_R - m_t \bar c_R t_L ) H \,, \notag \\
-i \bar t_R \gm c_R \pam H & \to & (m_c \bar t_R c_L - m_t \bar t_L c_R) H \,.
\end{eqnarray}
These substitutions are exactly the ones that can be applied for on-shell $t$ and $c$ quarks, because $\pam H = i q_\mu H$, with $q = p_t - p_c$, and the Dirac equation can be used to simplify $\ptsl$ and $\pcsl$. However, the operator equality in Eq.~(\ref{ec:Oeq3}) is more general, and valid for off-shell particles as well, provided that all contributions from gauge-invariant operators are taken into account.

\section{Rewriting derivative terms in amplitude calculations}
\label{sec:b}

\begin{figure}[htb]
\begin{center}
\begin{tabular}{ccc}
\epsfig{file=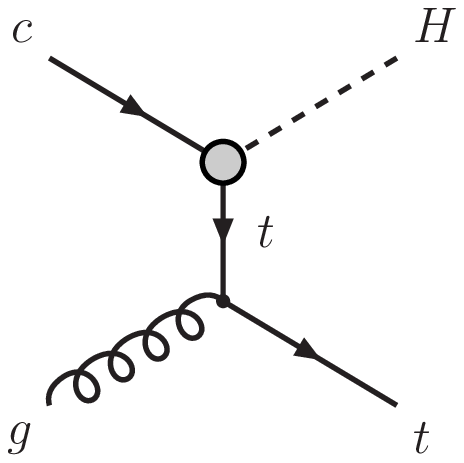,height=3cm,clip=} & \quad \quad &
\epsfig{file=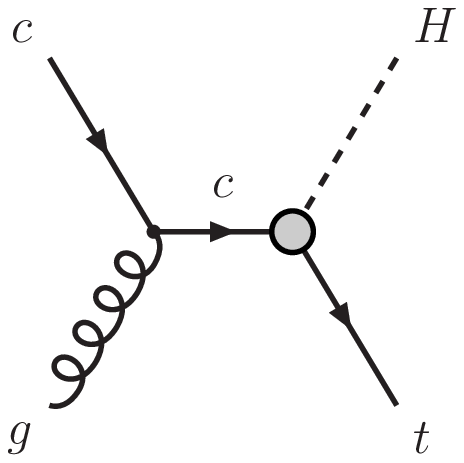,height=3cm,clip=}
\end{tabular}
\caption{Feynman diagrams contributing to $gc \to tH$.}
\label{fig:diag-tH}
\end{center}
\end{figure}

As we have seen in appendix~\ref{sec:a},
the operator equalities in Eqs.~(\ref{ec:Oeq}) imply that derivative terms in $Htt$, $Htq$ interactions can be replaced in amplitude calculations by scalar and pseudo-scalar terms, if all contributions from the gauge-invariant effective operators involved are taken into account in the amplitudes and a gauge-invariant set of Feynman diagrams is considered. For example, the 
replacements for the $Htc$ vertex are
\begin{eqnarray}
\bar c_L \qsl t_L & \to & m_t \bar c_L t_R - m_c \bar c_R t_L \,, \notag \\
\bar c_R \qsl t_R & \to & m_t \bar c_R t_L - m_c \bar c_L t_R \,,
\label{ec:Orepl}
\end{eqnarray}
plus the hermitian conjugate, being $q=p_t-p_c$ the (outgoing) Higgs boson momentum.
These replacements are trivial if both the top and charm quarks are on-shell, because in this case they follow from the application of the Dirac equation. In this appendix we show explicitly with a simple amplitude calculation that this is also the case for off-shell quarks. For convenience, we define the ``off-shell'' operator
\begin{equation}
\mathcal{O}_\text{off} = \bar c \left[ \qsl (\FL P_L + \FR P_R) - m_t (\FR P_L + \FL P_R)
+ m_c (\FL P_L + \FR P_R) \right] t + \text{H.c.} \,,
\end{equation}
where $\FL$ and $\FR$ are complex constants.
This operator is defined as an arbitrary derivative contribution minus the (pseudo-)scalar terms by which it can be replaced, according to Eqs.~(\ref{ec:Orepl}). It can also be written as
\begin{equation}
\mathcal{O}_\text{off} = (\FR P_L + \FL P_R) (\ptsl - m_t) - (\pcsl - m_c)  (\FL P_L + \FR P_R) 
+ \text{H.c.} \,,
\end{equation}
which is a much more convenient form for computations. Here
we calculate the amplitude for $gc \to tH$, involving the two diagrams in Fig.~\ref{fig:diag-tH}.
The gray circles stand for an anomalous vertex involving $\mathcal{O}_\text{off}$. In order to show that the replacements
in Eqs.~(\ref{ec:Orepl}) can be done in this process (where the $\mathcal{O}_\text{off}$ anomalous interaction involves off-shell quarks) it is sufficient to show that the sum of the amplitudes corresponding to both diagrams cancel. In this simple process, the only contributions to the amplitude from the gauge-invariant operators considered are the $Htc$ trilinear vertices, although this is not always the case, and for gauge boson vertices additional quartic diagrams sometimes appear~\cite{AguilarSaavedra:2008zc,AguilarSaavedra:2008gt}.

We label the momenta of the gluon, the external charm and top quarks and the Higgs boson as $p_1$, $p_2$, $p_3$ and $p_4$, respectively. Using the on-shell conditions for the charm (top) quark
involved in the anomalous interaction in the first (second) diagram and using
$(\psl + m) (\psl - m) = p^2-m^2$ to simplify the propagators,
the two amplitudes read, in standard notation,
\begin{eqnarray}
\mathcal{M}_t & = & - g_s \frac{\lambda^a}{2} \bar u(p_3) \gm (\FL P_L + \FR P_R) u(p_2) \; \varepsilon_\mu(p_1) \,, \notag \\
\mathcal{M}_s & = & g_s \frac{\lambda^a}{2} \bar u(p_3) (\FR P_L + \FL P_R) \gm  u(p_2) \; \varepsilon_\mu(p_1) \,,
\end{eqnarray}
and it is clear that $\mathcal{M}_t + \mathcal{M}_s = 0$.

\section{Operator contributions to top-Higgs interactions and quark masses}
\label{sec:c}

We collect here the effective operator contributions to top-Higgs interactions and quark masses. We use the shorthand $\alpha_x = C_x/\Lambda^2$, and drop the indices on the $\alpha$ constants. Our expressions coincide with those in Refs.~\cite{Whisnant:1997qu,Yang:1997iv}.
The contribution of the operators in Eqs.~(\ref{ec:Oall}) with $i=j=3$ to the $Htt$ vertex are
\begin{eqnarray}
\alpha O_{\phi q}^{(3,33)} + \alpha^* (O_{\phi q}^{(3,33)})^\dagger & \supset &
\IM \alpha \; v \,  \bar t_L \gm t_L \, \pam H
  \,, \notag \\
\alpha O_{\phi q}^{(1,33)} + \alpha^* (O_{\phi q}^{(1,33)})^\dagger & \supset &
- \IM \alpha \; v \, \bar t_L \gm t_L \, \pam H
  \,, \notag \\
\alpha O_{\phi u}^{33} + \alpha^* (O_{\phi u}^{33})^\dagger & \supset &
- \IM \alpha \; v \, \bar t_R \gm t_R \, \pam H
  \,, \notag \\
\alpha O_{u\phi}^{33} + \alpha^* (O_{u\phi}^{33})^\dagger & \supset &
\frac{3v^2}{2\sqrt 2} \left[ \RE \alpha \;  \bar t \; t
+ i \, \IM \alpha \; \bar t \gamma_5 t \right]  H
  \,.
\end{eqnarray}
The contributions to top FCN vertices with the Higgs boson are given by
\begin{eqnarray}
\alpha O_{\phi q}^{(3,ij)} + \alpha^* (O_{\phi q}^{(3,ij)})^\dagger & \supset &
-i \, \frac{v}{2} \left[ \alpha \, \bar u_{Li} \gm u_{Lj} - \alpha^* \bar u_{Lj} \gm u_{Li} \right] \pam H
  \,, \notag \\
\alpha O_{\phi q}^{(1,ij)} + \alpha^* (O_{\phi q}^{(1,ij)})^\dagger & \supset &
i \, \frac{v}{2} \left[ \alpha \, \bar u_{Li} \gm u_{Lj} - \alpha^* \bar u_{Lj} \gm u_{Li} \right] \pam H
  \,, \notag \\
\alpha O_{\phi u}^{ij} + \alpha^* (O_{\phi u}^{ij})^\dagger & \supset &
i \, \frac{v}{2} \left[ \alpha \, \bar u_{Ri} \gm u_{Rj} - \alpha^* \bar u_{Rj} \gm u_{Ri} \right] \pam H
  \,, \notag \\
\alpha O_{u\phi}^{ij} + \alpha^* (O_{u\phi}^{ij})^\dagger & \supset &
\frac{3v^2}{2\sqrt 2} \left[ \alpha \, \bar u_{Li} u_{Rj} 
+ \alpha^* \bar u_{Rj} u_{Li} \right] H
\,,
\end{eqnarray}
with $i,j=1,3/3,1$ for $Htu$ and $i,j=2,3/3,2$ for $Htc$. Finally, the only operator contributing to quark masses is $O_{u\phi}^{ij}$,
\begin{equation}
\alpha O_{u\phi}^{ij} + \alpha^* (O_{u\phi}^{ij})^\dagger \supset
\frac{v^3}{2\sqrt 2} \left[ \alpha \, \bar u_{Li} u_{Rj} 
+ \alpha^* \bar u_{Rj} u_{Li} \right] \,.
\end{equation}


\begin{thebibliography}{99}

\bibitem{Burges:1983zg}
  C.~J.~C.~Burgess and H.~J.~Schnitzer,
  Nucl.\ Phys.\  B {\bf 228} (1983) 464

\bibitem{Leung:1984ni}
  C.~N.~Leung, S.~T.~Love and S.~Rao,
  Z.\ Phys.\  C {\bf 31} (1986) 433

\bibitem{Buchmuller:1985jz}
  W.~Buchmuller and D.~Wyler,
  Nucl.\ Phys.\  B {\bf 268} (1986) 621

\bibitem{Gounaris:1996vn}
  G.~J.~Gounaris, M.~Kuroda and F.~M.~Renard,
  Phys.\ Rev.\  D {\bf 54} (1996) 6861
  {\tt [hep-ph/9606435]}

\bibitem{Gounaris:1996yp}
  G.~J.~Gounaris, D.~T.~Papadamou and F.~M.~Renard,
  Z.\ Phys.\  C {\bf 76} (1997) 333
  {\tt [hep-ph/9609437]}

\bibitem{Whisnant:1997qu}
  K.~Whisnant, J.~M.~Yang, B.~L.~Young and X.~Zhang,
  Phys.\ Rev.\  D {\bf 56} (1997) 467
  {\tt [hep-ph/9702305]}

\bibitem{Yang:1997iv}
  J.~M.~Yang and B.~L.~Young,
  Phys.\ Rev.\  D {\bf 56} (1997) 5907
  {\tt [hep-ph/9703463]}

\bibitem{delAguila:2000rc}
  F.~del Aguila, M.~Perez-Victoria and J.~Santiago,
  JHEP {\bf 0009} (2000) 011
  {\tt [hep-ph/0007316]}

\bibitem{delAguila:2000aa}
  F.~del Aguila, M.~Perez-Victoria and J.~Santiago,
  Phys.\ Lett.\  B {\bf 492} (2000) 98
  {\tt [hep-ph/0007160]}

\bibitem{Ferreira:2008cj}
  P.~M.~Ferreira, R.~B.~Guedes and R.~Santos,
  Phys.\ Rev.\  D {\bf 77} (2008) 114008
  {\tt [0802.2075 [hep-ph]]}

\bibitem{Coimbra:2008qp}
  R.~A.~Coimbra, P.~M.~Ferreira, R.~B.~Guedes, O.~Oliveira, A.~Onofre, R.~Santos and M.~Won,
  Phys.\ Rev.\  D {\bf 79} (2009) 014006
  {\tt [0811.1743 [hep-ph]]}

\bibitem{Ferreira:2009bf}
  P.~M.~Ferreira and R.~Santos,
  {\tt 0903.4470 [hep-ph]}

\bibitem{AguilarSaavedra:2008zc}
  J.~A.~Aguilar-Saavedra,
  Nucl.\ Phys.\  B {\bf 812} (2009) 181
  {\tt [0811.3842 [hep-ph]]}

\bibitem{Georgi:1991ch}
  H.~Georgi,
  Nucl.\ Phys.\  B {\bf 361} (1991) 339

\bibitem{Arzt:1993gz}
  C.~Arzt,
  Phys.\ Lett.\  B {\bf 342} (1995) 189
  {\tt [hep-ph/9304230]}

\bibitem{Grzadkowski:2003tf}
  B.~Grzadkowski, Z.~Hioki, K.~Ohkuma and J.~Wudka,
  Nucl.\ Phys.\  B {\bf 689} (2004) 108
  {\tt [hep-ph/0310159]}


\bibitem{AguilarSaavedra:2008gt}
  J.~A.~Aguilar-Saavedra,
  Nucl.\ Phys.\  B {\bf 804} (2008) 160
  {\tt [0803.3810 [hep-ph]]}

\bibitem{AguilarSaavedra:2006fy}
  J.~A.~Aguilar-Saavedra, J.~Carvalho, N.~Castro,  A.~Onofre and F.~Veloso,
  Eur.\ Phys.\ J.\  C {\bf 50} (2007) 519
  {\tt [hep-ph/0605190]}

\bibitem{AguilarSaavedra:2000aj}
  J.~A.~Aguilar-Saavedra and G.~C.~Branco,
  Phys.\ Lett.\  B {\bf 495} (2000) 347
  {\tt [hep-ph/0004190]}

\end{thebibliography}
\end{document}